\documentclass[11pt,a4paper]{article}
\pdfoutput=1 

\usepackage[table]{xcolor}
\usepackage{colortbl}
\RequirePackage{ifpdf} 
\usepackage{amsmath} 
\usepackage{mathtools}
\usepackage{subfigure}
\usepackage{multirow}

\usepackage{jheppub}
\usepackage{pstricks}
\usepackage[final]{pdfpages} 
\usepackage{ifpdf} 
\usepackage{slashed}

\usepackage[normalem]{ulem}
\usepackage{color}
\usepackage{xcolor}
\definecolor{urlblue}{rgb}{0.2,0.4,0.7}
\definecolor{citegreen}{rgb}{0,0.6,0.2}
\definecolor{linkred}{rgb}{0.9,0.2,0.1}
\usepackage{hyperref}
\hypersetup{
colorlinks=true, citecolor=citegreen, linkcolor=blue, urlcolor=urlblue}

\usepackage{graphics}
\usepackage{etoolbox} 
\usepackage{fixmath}
\usepackage{psfrag}
\usepackage{diagbox}

\usepackage{notoccite} 

\usepackage{amsfonts}
\usepackage{autobreak}
\usepackage{marginnote}
\usepackage{enumitem}
\usepackage{appendix}
\usepackage{caption} 
\newcommand{\NOdisplay}[1]{ }

\newcommand{\afb}{A_{\rm FB}}
\newcommand{\afblr}{A_{\rm pol,FB}}

\newcommand{\sia}{\sigma_A}
\newcommand{\sis}{\sigma_S}
\newcommand{\as}{\alpha_s}

\newcommand{\tbart}{t{\bar t}}
\newcommand{\bbarb}{b{\bar b}}
\newcommand{\QbarQ}{Q{\bar Q}}

 \newcommand{\qm}{\boldsymbol{p}_1}
 \newcommand{\qp}{\boldsymbol{p}_2}
 \newcommand{\km}{\boldsymbol{k}_1}
 \newcommand{\kp}{\boldsymbol{k}_2}
 \newcommand{\ki}{\boldsymbol{k}_i}
 \newcommand{\nt}{\boldsymbol{n}_T}
 
\newcommand{\bc} {\begin{center}}
\newcommand{\ec} {\end{center}}

\newcommand{\simgt}{\hbox{ \raise3pt\hbox to 0pt{$>$}
    \raise-3pt\hbox{$\sim$} }}
\newcommand{\simsm}{\hbox{ \raise3pt\hbox to 0pt{$<$}
    \raise-3pt\hbox{$\sim$} }}
    
\title{Top and bottom quark forward-backward asymmetries at next-to-next-to-leading order QCD in 
    (un)polarized electron positron collisions}

\author[a]{Werner Bernreuther,}
\author[b]{Long Chen,}
\author[b]{Peng-Cheng Lu,}
\author[b]{Zong-Guo Si}

\affiliation[a]{Institut f\"ur Theoretische Teilchenphysik und Kosmologie, \\
    RWTH Aachen University,  52056 Aachen, Germany}
\affiliation[b]{School of Physics, Shandong University, Jinan, Shandong 250100, China}

\emailAdd{breuther@physik.rwth-aachen.de}
\emailAdd{longchen@sdu.edu.cn}
\emailAdd{pclu@sdu.edu.cn}
\emailAdd{zgsi@sdu.edu.cn}

\preprint{}

\abstract{
We consider, at order $\as^2$ in the QCD coupling,  top-quark pair production in the continuum at various center-of-mass energies and
    $b$-quark pair production at the $Z$ resonance by (un)polarized electron and positron beams. For top quarks we compute the forward-backward asymmetry
      with respect to the top-quark direction of flight, the associated polar angle distribution, and we analyze the effect of 
      beam polarization on the QCD corrections to the leading-order asymmetry. We calculate also the polarized forward-backward asymmetry.
     For $b$-quark production at the $Z$ peak we explore different definitions of $\afb$. In particular,
      we analyze $b$ jets defined by  the Durham and  the flavor-$k_T$ clustering algorithms.
      We compute the  inclusive $b$-jet and two-jet asymmetry with respect to the  $b$-jet direction. For the latter asymmetry the QCD corrections 
      to order $\as^2$ are small. That predestines it to act as a precision observable.
}

\begin{document}
\allowdisplaybreaks[4]
\unitlength1cm
\maketitle
\flushbottom

\section{Introduction} 
\label{sec:intro}
Among the various high-energy particle accelerator types that are presently discussed  for potential
future realization, an electron-positron collider has the highest priority. Several proposals for a linear
\cite{ILCInternationalDevelopmentTeam:2022izu,Brunner:2022usy,Dasu:2022nux} and a circular \cite{Bernardi:2022hny,Gao:2022lew}
$e^+e^-$ collider have been made. The major motivation for such a machine is to construct a Higgs factory with which the properties
of the 125 GeV Higgs boson can be studied with unprecedented precision. Besides, such a machine would allow for new physics searches and 
 precision studies of weak gauge bosons and heavy quarks, in particular top quarks (see, for instance, \cite{Blondel:2019ykp,CEPCPhysicsStudyGroup:2022uwl}).
 Longitudinal polarization of $e^-$ and $e^+$ beams, which is a prospective option especially for linear colliders, is a further asset in the exploration of the fundamental interactions
  at high energies. 

As far as the physics of top and bottom quarks at a future linear or circular $e^+ e^-$ collider is concerned, the measurement of forward-backward asymmetries $(\afb)$ will play a prominent role, 
in particular for the precision determinations of electroweak parameters.
On the theory side, this requires precise predictions, in particular of the higher-order QCD corrections to these asymmetries.

For massive quark-antiquark production in $e^+e^-$ collisions the following Standard Model (SM) radiative corrections to the lowest-order forward-backward asymmetry are known so far.
The fully massive next-to-leading order (NLO) electroweak and QCD corrections were determined in \cite{Beenakker:1991ca,Bohm:1989pb,Bardin:1999yd,Freitas:2004mn,Fleischer:2003kk} and 
in \cite{Jersak:1981sp,Arbuzov,Djouadi:1994wt,Hahn:2003ab,Khiem:2012bp,Arbuzov:2020ghr}, respectively.  
The heavy quark pair production cross section was computed to order $\alpha_s^2$ and order $\alpha_s^3$ in \cite{Gorishnii:1986pz,Chetyrkin:1996cf,Chetyrkin:1997qi,Chetyrkin:1997pn} and \cite{Hoang:2008qy,Kiyo:2009gb}, respectively, 
using approximations.\footnote{The complete electroweak two-loop corrections for typical precision observables at the Z resonance were finalized 
in \cite{Dubovyk:2018rlg}; cf. also further references therein.}~
The fully differential $\tbart$ cross section at NLO QCD with $t\to Wb$ decay was investigated in \cite{Bach:2017ggt}
and recently, the NLO QCD corrections to off-shell $\tbart$ production with semi-leptonic top-quarks decays were computed in \cite{Denner:2023grl}.
The full next-to-next-to-leading order (NNLO), i.e. the order $\as^2$ QCD corrections to $\afb$ 
 were computed by \cite{Gao:2014eea,Chen:2016zbz} for the top quark in $t{\bar t}$ production above the production threshold. 
A considerable effort was made to investigate $t{\bar t}$ production at threshold, presently known at next-to-next-to-next-to-leading (NNNLO) order QCD~\cite{Beneke:2015kwa,Beneke:2016cbu,Beneke:2016kkb,Beneke:2017rdn}.
 Recently, the NNNLO QCD correction to $e^+e^- \rightarrow t{\bar t}$ production in the continuum via a virtual photon was obtained in 
\cite{Chen:2022vzo} with the powerful auxiliary-mass flow method \cite{Liu:2017jxz,Liu:2020kpc,Liu:2021wks,Liu:2022mfb,Liu:2022chg}.
For $\bbarb$ production at the $Z$ peak, the order $\as^2$ corrections were calculated for several definitions of the $b$-quark forward-backward (FB) asymmetry for massless $b$ quarks by \cite{Altarelli:1992fs,Ravindran:1998jw,Catani:1999nf,Weinzierl:2006yt} and for massive quarks in \cite{Bernreuther:2016ccf}, and also in \cite{Wang:2020ell} where the optimization based on the principle of maximum conformality \cite{Brodsky:2012rj,Brodsky:2011ig,Mojaza:2012mf} was taken into account.

In this paper we extend the present theory knowledge of the top- and bottom-quark $\afb$ in several ways. In particular we investigate the effect of beam polarization on these asymmetries. 
As far as $\tbart$ production in the continuum is concerned we calculate the order $\as^2$ QCD corrections to the top-quark $\afb$ with respect to the top-quark direction of flight 
for a set of longitudinal polarizations of the $e^-$ and $e^+$ beams and compare with the respective $\afb$ resulting from top-quark production with unpolarized beams. 
We find a marginal dependence of the QCD corrections on  beam polarization.
As to $e^+ e^- \to \bbarb$  at the $Z$ resonance, we consider also production by polarized beams and compare with results for unpolarized beams.
We compute the order $\as^2$ QCD corrections to the $b$-quark $\afb$ for the cases where the forward and backward hemispheres are defined with respect to the $b$-quark direction of flight and the oriented thrust axis. 
Moreover, we consider $b$ jets defined by the Durham \cite{Catani:1991hj}
and flavor-$k_T$ \cite{Banfi:2006hf} algorithm, use the resulting  $b$-jet axis for defining $\afb$, and compute the order $\as^2$ QCD corrections to an inclusive $b$-jet asymmetry and to the two-jet asymmetry. 
The latter $\afb$ was calculated before to order $\as$  in \cite{Djouadi:1994wt} and for massless $b$ quarks to order $\as^2$ in \cite{Weinzierl:2006yt}. 
As in the massless case, the QCD corrections to this asymmetry are small. Thus it may serve as a suitable precision observable. 
     
Our paper is organized as follows. section~\ref{sec:FB} contains formulas for $\afb$ to order $\as^2$, unexpanded and expanded in the 
    QCD coupling, for the polarized forward-backward asymmetry \cite{Blondel:1987gp}, and a set of beam polarizations that will be used in the following sections.
    In section~\ref{sec:asyt} we investigate $\tbart$ production to order $\as^2$ by polarized and unpolarized beams at c.m. energies 380~GeV, 400~GeV, 500~GeV, and 700~GeV.
    The forward and backward hemispheres are defined with respect to the top-quark direction of flight and we compute the symmetric $(\sigma_S)$ and antisymmetric $(\sigma_A)$ 
    cross sections and $\afb$ to order $\as^2$. We analyze the effect of beam polarization on the QCD corrections to the leading-order FB asymmetry. 
    For the c.m. energy 500~GeV we determine the polar angle distribution of the top quark for (un)polarized beams. In addition we calculate the polarized FB asymmetry
    for the c.m. energies listed above. In section~\ref{sec:asyb} we consider $\bbarb$ production to order $\as^2$ at the $Z$-boson resonance by (un)polarized beams.
    Here we explore different definitions of the forward and backward hemispheres. First we use the $b$-quark axis and the oriented thrust axis
     and determine $\sigma_S$, $\sigma_A$, and the resulting $\afb$. Then we turn to $b$ jets defined by i) the Durham and ii) the flavor-$k_T$ clustering algorithm.
     Related to the respective $b$-jet direction we compute the  inclusive $b$-jet and two-jet asymmetry.
     We comment also briefly on the electroweak corrections to the $b$-qaurk asymmetries.
    We conclude in section~\ref{sec:concl}.

\section{The forward-backward and polarized forward-backward asymmetry}
\label{sec:FB}
In this paper we consider the production of top-quark and bottom-quark pairs  in (un)polarized $e^+ e^-$ collisions,
\begin{align} \label{eeQQX}
e^-(\qm) +  e^+(\qp)  \rightarrow Q(\km)  + \bar{Q}(\kp) + X \, , \quad Q= t, b \, ,
\end{align} 
to lowest order in the electroweak couplings and to second order in the QCD coupling $\as$. 
The three-momenta displayed in \eqref{eeQQX} refer to the $e^+e^-$ c.m. frame.
 Top-quark production is analyzed in the continuum away from the $\tbart$ threshold where perturbation theory is applicable. In the case of $Q=b$ we confine
ourselves to $\bbarb$ production at the $Z$ resonance. 

The differential cross section to order $\as^2$ of the reaction \eqref{eeQQX} was computed in \cite{Chen:2016zbz} using the antenna subtraction framework. 
We extend these results to $\tbart$ and $\bbarb$ production with polarized beams. 
A brief outline of the computational details is given at the end of this section.

We consider massless electrons. 
 As usual the $e^-$ and $e^+$ beam polarizations are described by the following polarizations 
 projectors for the $u$ and $v$ spinors:
 \begin{eqnarray} \label{eq:beamPOL}
 u(p_1,P_L) \otimes \bar{u}(p_1,P_L) &=& \slashed{p}_1 \big( 1 + P_L\, \gamma_5 \big) \, ,\nonumber\\
 v(p_2,P_R) \otimes \bar{v}(p_2,P_R) &=& \slashed{p}_2 \big( 1 + P_R\, \gamma_5 \big)\, ,
\end{eqnarray}
 where $P_L$ is the left-handed polarization of the electron 
(+1 = fully left-handed, 0 = unpolarized, -1 = fully right-handed) and $P_R$ is the right-handed polarization of the positron 
($+1, 0, -1$ is fully-right-handed, unpolarized, and fully left-handed, respectively).
Furthermore, we define
\begin{equation} \label{defPvPa}
       P_v = 1+ P_L P_R \, , \quad P_a = P_L + P_R
\end{equation}
and for notational clarity, we use the notation 
 $e^-_{L} \equiv P_L$ and $e^+_{R} \equiv P_R$ in the following.
 For  unpolarized beams, one has $P_v =1\,, P_a = 0$.
 In the computations of the next sections we 
 consider four benchmark polarization configurations listed in table~\ref{tab:benchpol}.

\vspace{2mm}
\begin{table}[htbp]
\begin{center}
  \caption{Benchmark polarizations of the $e^-$ and $e^+$ beams used in the computations below and corresponding values of the 
    polarization combinations \eqref{defPvPa}. } 
  \vspace{1mm}
  {\renewcommand{\arraystretch}{1.2}
\renewcommand{\tabcolsep}{0.2cm}
\begin{tabular*}{0.40 \textwidth}{|@{\extracolsep{\fill} } c| c| c| c|   } \hline  
  $e^{-}_L$ & $e^{+}_R$ & $P_v$ & $P_a$ \\ \hline\vspace{0.4mm}
  $-80\% $ & $+30\%$ & $0.76$ & $-0.5$ \\
   $+80 \%$ & $-30\%$ & $0.76$ & $~~0.5$ \\
   $+80\%$ & $+30\%$ & $1.24$ & $~~1.1$ \\
   $-80\%$ & $-30\%$ & $1.24$ & $-1.1$\\ \hline 
   \end{tabular*} }
\label{tab:benchpol}
\end{center}
\end{table}

The forward-backward asymmetry $A_{\rm FB}$ for the production of a massive quark $Q$ is defined by
\begin{eqnarray}
A_{\rm FB} \equiv \frac{N_F - N_B}{N_F + N_B} \, ,
\label{AFBdef}
\end{eqnarray}
where $N_F$ $(N_B)$ is the number of quarks $Q$ produced in the forward (backward) direction. 
The identification of the forward and backward direction involves a choice of reference axis. 
The definition of the reference axis must be such that the resulting forward-backward asymmetry 
is an infrared safe (IR-safe) quantity so that it can be reliably calculated. 
The asymmetry $A_{\rm FB}$ is generated by those terms in the squared S-matrix elements of the reaction \eqref{eeQQX}
 that are odd under the interchange of $Q$ and $\bar Q$ (while the initial state is kept fixed).

The asymmetry $A_{\rm FB}$ can also be expressed
conveniently in terms of the symmetric and antisymmetric cross section $\sigma_S$ and $\sigma_A$ 
for the inclusive production of the heavy quark $Q$, i.e.,
\begin{eqnarray} \label{defafb}
A_{\rm FB}=\frac{\sigma_A}{\sigma_S} =\frac{\sigma_F-\sigma_B}{\sigma_F+\sigma_B}.
\end{eqnarray}
The $\sigma_F$ and $\sigma_B$ are the forward and backward cross sections, respectively.

To order $\alpha_s^2$ the symmetric and antisymmetric cross sections receive the following perturbative contributions:
\begin{align}\label{sasbeitraege}
\sigma_{A,S} = \sigma_{A,S}^{(2,0)} +\sigma_{A,S}^{(2,1)} + \sigma_{A,S}^{(3,1)} + \sigma_{A,S}^{(2,2)} 
+ \sigma_{A,S}^{(3,2)} + \sigma_{A,S}^{(4,2)} + \mathcal{O}(\as^3) \, ,
\end{align}
where the first number in the superscripts $(i,j)$ denotes the number of final-state partons associated with the respective  term and the second one the order of $\alpha_s$. 
Inserting (\ref{sasbeitraege}) into (\ref{defafb}) we get the {\it{unexpanded}} $A_{\rm FB}$ to first and to second order in $\as$:
\begin{align}
\afb(\as) = & \frac{\sigma_{A}^{(2,0)} +\sigma_{A}^{(2,1)}  + \sigma_{A}^{(3,1)}}{{\sigma_{S}^{(2,0)} 
 +\sigma_{S}^{(2,1)} + \sigma_{S}^{(3,1)}}} \equiv  \afb^{\rm LO}~C_1\, , \label{afb1unex} \\
\afb(\as^2) = & \frac{\sigma_{A}^{(2,0)} +\sigma_{A}^{(2,1)}  + \sigma_{A}^{(3,1)} + \sigma_{A}^{(2,2)}+ \sigma_{A}^{(3,2)} +
\sigma_{A}^{(4,2)}}{\sigma_{S}^{(2,0)} +\sigma_{S}^{(2,1)} + \sigma_{S}^{(3,1)}  + \sigma_{S}^{(2,2)} + \sigma_{S}^{(3,2)} +
\sigma_{S}^{(4,2)}} \equiv  \afb^{\rm LO}~C_2\, , \label{afbgesamt}
\end{align}
where 
\begin{equation}
 \afb^{\rm LO} = \frac{\sia^{(2,0)}}{\sis^{(2,0)}} 
\label{eq:afb0}
\end{equation}
is the forward-backward asymmetry at Born level. 
The factors $C_1$ and $C_2$, defined by the respective ratio on the left-hand side of eq.~\eqref{afb1unex} and~\eqref{afbgesamt}, are the unexpanded first- 
and second-order QCD correction factors.

Taylor expanding eq.~\eqref{afb1unex} to first order and eq.~\eqref{afbgesamt} to second order in  $\as$ yields the {\it{expanded}} $A_{\rm FB}$: 
\begin{align}
\afb^{\rm NLO}= & \afb^{\rm LO} \left[1+A_1\right] \;+\; {\cal O}(\as^2) \, , \label{afb1ent} \\
\afb^{\rm NNLO} =&
\afb^{\rm LO}\;\left[1\;+\;A_{1} \;+\;
A_{2}\right]   \;+\; {\cal O}(\as^3)     \, , \label{afbent}
\end{align}
where $A_{1}$ and $A_{2}$ are the QCD corrections of ${\cal O}(\as)$ and  ${\cal O}(\as^2)$, respectively.
\begin{align}
A_{1} =& \sum\limits_{i=2,3} \big[\frac{\sia^{(i,1)}}{\sia^{(2,0)}} \;-\;\frac{\sis^{(i,1)}}{\sis^{(2,0)}} \big] \, ,\label{eq:afb1} \\
A_{2} =& \sum\limits_{i=2,3,4} \big[ \frac{\sia^{(i,2)}}{\sia^{(2,0)}} \;-\;
\frac{\sis^{(i,2)}}{\sis^{(2,0)}} \big]
 - \frac{\sis^{(2,1)}+\sis^{(3,1)}}{\sis^{(2,0)}} \, A_1  \, .\label{eq:afb2}
\end{align}
eqs.~\eqref{afb1ent} and~\eqref{afbent} are the expanded forms of the forward-backward asymmetry at NLO and NNLO QCD.
The unexpanded and expanded first- and second-order forward-backward asymmetries differ by terms of order $\as^2$ and order $\as^3$, respectively. 
The differences between the two forms may be considered as an estimate of the theory uncertainties.
~\\

With polarized beams, one can consider also the so-called polarized forward-backward asymmetry \cite{Blondel:1987gp}:
\begin{equation}
\afblr = \frac{1}{P} \frac{(\sigma_F(P) - \sigma_F(-P) ) - (\sigma_B(P)  - \sigma_B(-P) )}{(\sigma_F(P) + \sigma_F(-P) ) + (\sigma_B(P)  + \sigma_B(-P) )} \, , 
\label{defalrfb}
\end{equation}
where $P \equiv P_a/P_v$ (sometimes called the polarization degree of the $e^+e^-$ system) with $P_a,\, P_v$ defined in eqs.(\ref{defPvPa}) and (\ref{eq:beamPOL}), and $\sigma_{F,B}(\pm P)$  is the heavy-quark cross section with beam polarization $\pm P$ 
in the forward hemisphere ($0 \leq  \cos \theta \leq 1$) and backward hemisphere ($-1 \leq \cos \theta \leq 0$), respectively. In the case of $\tbart$
 production $\theta$ is the angle between the electron and the top quark. In the case of $\bbarb$ production at the $Z$ peak, $\theta$ is the angle between
  the electron and the  axis that defines the forward direction, cf. section~\ref{sec:asyb}.
The variable $\afblr$ combines data from different polarization configurations. 
We will see that, to lowest order in the electroweak couplings, the QCD correction factors to this observable remain 
exactly the same as in the unpolarized case. Unexpanded and expanded versions of this asymmetry can be obtained in complete analogy to 
\eqref{afb1unex}, \eqref{afbgesamt} and \eqref{afb1ent}, \eqref{afbent}, respectively, and we calculate both of them below.

In our computations we use for the on-shell top and $b$ mass and other parameters \cite{Workman:2022ynf} (we use the $G_\mu$ scheme):
\begin{eqnarray} 
m_t = 172.5 ~\rm GeV\, , & m_b = 4.78 ~\rm GeV \, , &  m_Z = 91.1876 ~\rm GeV\, ,\nonumber\\
\sin^2\theta_W = 0.2229\, , & \alpha_s(m_Z) = 0.11805 \, , & \alpha_{em} = 0.00756 \,. 
\label{eq:parameter}
\end{eqnarray}

For completeness, we sketch here our computational set-up that is a straightforward extension to polarized beams of the approach developed in \cite{Chen:2016zbz,Bernreuther:2016ccf}.
To the order of perturbation theory we are working the cross section of the reaction \eqref{eeQQX} receives contributions from the two-parton $\QbarQ$ state 
(at Born level, to order $\as$, and to order $\as^2$), the three-parton state $\QbarQ g$ (to order $\as$ and to order $\as^2$), and the four-parton states 
 $\QbarQ gg$, $\QbarQ q{\bar q}$, and above the $4Q$ threshold from $\QbarQ \QbarQ$ (to order $\as^2$). In the case of $\tbart$ production,
 all quarks $q$ besides the top quark are taken to be massless, while in the case of $\bbarb$ production at the $Z$ peak, the $b$-quark mass is taken to be massive.
 
As to the renormalization procedure used, as usual the QCD coupling $\as(\mu)$ is defined in the ${\overline{\rm MS}}$ scheme at the chosen renormalization scale $\mu$ while the mass of the heavy quark is defined in the on-shell scheme.

 We work to lowest order in the electroweak couplings. Thus, each of the just-mentioned various contributions $d\sigma^{(i,j)}$ 
   to the differential $b {\bar b}$ cross section to
  order $\alpha_s^2$  is given, at arbitrary 
   c.m. energy, by the sum of an s-channel $\gamma$ and $Z$-boson contribution  and a $\gamma Z$ interference term. 
   The $d\sigma^{(i,j)}$ are of the form
   \begin{equation}\label{eq:strucsig} 
   d\sigma^{(i,j)} = \sum\limits_{a=\gamma, Z, \gamma Z} F_a^{(j)} \: L^{\mu\nu}_a H^{(i,j)}_{a,\mu\nu} \: d\Phi_i \, .
   \end{equation}
   The first index $i$ in the superscript $(i,j)$ labels the final state, i.e., $i= b{\bar b}$, $b{\bar b}g$, $b{\bar b}gg$, $b{\bar b}q {\bar q}$ $(q=u,d,s,c,b)$.
Here  $d\Phi_i$ denotes the $i$-particle phase-space measure and $L^{\mu\nu}_a$ are the lepton tensors (with the boson propagators included)
that contain, as compared to unpolarized beams, additional terms due to the polarization projectors   \eqref{eq:beamPOL}. 
The tensors $H^{(i,j)}_{a,\mu\nu}$ are the antenna-subtracted, i.e., infrared finite parton tensors of order $\alpha_s^j$  \cite{Chen:2016zbz}.
The antenna subtraction terms that remove the soft and collinear divergences to order $\as^2$  from the $Q{\bar Q}gg$ and $Q{\bar Q}q{\bar q}$
 matrix elements were constructed in \cite{Bernreuther:2013uma} and \cite{Bernreuther:2011jt}, respectively, while the subtraction terms for the  $Q{\bar Q}g$ final state to order $\as^2$ were
  determined in~\cite{Dekkers:2014hna}.
 Thus the $d\sigma^{(i,j)}$ are finite by construction and, therefore,  the Lorentz contractions and the phase-space integration in \eqref{eq:strucsig} can be done in $D=4$ dimensions.
 The factors  $F_a^{(j)}$ contain the electroweak couplings and the flux factor.
 

  Each contribution $(i,j)$ on the right-hand side of  \eqref{eq:strucsig} is separated into a  parity-even and -odd term. 
  To lowest order in the electroweak couplings these terms
  determine the cross sections $\sigma_S$ and $\sigma_A$ that are  symmetric and antisymmetric under the exchange of $Q$ and $\bar Q$, respectively.
   For the numerical evaluation of the $d\sigma^{(i,j)}$ we use the approach described in detail in  \cite{Chen:2016zbz}.
   
   In section~\ref{sec:asyb} we consider 
  $b{\bar b}$ production exactly at the $Z$ resonance. At this c.m. energy we neglect the s-channel $\gamma$ and $\gamma Z$ interference contributions to 
   the $d\sigma^{(i,j)}$ for determining the pure order $\as^2$ QCD corrections to various FB $b$-quark asymmetries. In addition, we briefly discuss also the impact of 
   the NLO electroweak corrections.

\section{Asymmetries for top quarks}
\label{sec:asyt}

We consider in this section $\tbart$ production to $\mathcal{O}(\alpha_s^2)$ for c.m. energies $380$~GeV, $400$~GeV, $500$~GeV, and  $700$~GeV.
The latter c.m. energy is above threshold for $4 t$ production. 
However, the $4 t$ final state, whose matrix element is ultraviolet and infrared finite at this order of perturbation theory, makes only a very small contribution to the cross section at this energy. 
Moreover, the $4 t$ production can be well separated experimentally from the other $\tbart +X$ final states. Therefore, we do not take this contribution into account in the results to be presented below. 
The various contributions to $\tbart$ production are conveniently classified as follows: flavor non-singlet (where the virtual $Z$ and $\gamma$ directly couple to the external $\tbart$, 
flavor singlet (where $\tbart$ is produced by a virtual gluon), and triangle or interference terms~\cite{Catani:1999nf,Bernreuther:2016ccf}.

We use the top-quark direction of flight in the $e^+e^-$ c.m. frame as reference axis  for defining the forward and backward hemisphere. 
This axis is infrared- and collinear-safe. The top-direction of flight can be reconstructed for instance with lepton plus jets events (or from all jets events)
 from $\tbart$ decay.

 Tables~\ref{ener21},~\ref{ener31},~\ref{ener11}, and~\ref{ener51} contain our results for the symmetric  $\tbart$ cross section, for the forward-backward asymmetry at LO, and the corrections  $A_1$ and $A_2$ 
  to $\afb^{\rm LO}$
  at NLO  and NNLO QCD in expanded form, both for unpolarized beams and the polarization configurations listed in table~\ref{tab:benchpol}, for the above c.m. energies. 
  The corresponding unexpanded NLO and NNLO QCD correction factors $C_1$ and $C_2$ are listed in table~\ref{c1c2}.
  The numbers for the QCD corrections given in these tables were obtained by setting the renormalization scale $\mu=\sqrt{s}$.
 The numbers in  super- and subscript correspond to the changes that result from setting the scale to $\mu=2\sqrt{s}$ and $\mu=\sqrt{s}/2$, respectively.
The results for the scale variations are derived by first obtaining the values of the symmetric and antisymmetric
cross sections $\sigma_S$, $\sigma_A$, respectively, at $\mu=2\sqrt{s}$ and $\mu=\sqrt{s}/2$ by means of the renormalization-group equation 
from which the scale uncertainties for the various quantities listed in the aforementioned tables are composed.
For the unexpanded $A_{\mathrm{FB}}$ and the corresponding $C_1, C_2$ coefficients being defined as ratios, it is expected that there is 
a cancellation between the scale dependence of the symmetric and antisymmetric cross sections if one chooses to vary both simultaneously.
Consequently, the scale uncertainties for these ratios are relatively small and may not display the usual improvement when the higher order perturbative corrections are included, 
see~table~\ref{c1c2}. 
Alternatively, one may choose to set the scale of $\sigma_S$ different from that of $\sigma_A$ to obtain a more conservative estimate for the scale uncertainties of these ratios. 
 However, we refrain from listing in these tables the scale uncertainties derived with these alternative conventions, both for the sake of not over-loading 
 the tables and also because their magnitudes are comparable to that of the scale uncertainties of $\sigma_S$, which are provided.
On the other hand, this feature is not exhibited in the QCD correction factors $A_1, A_2$ related to the expanded $A_{\mathrm{FB}}$, and 
one does observe the usual improvement when the higher order perturbative corrections are included, see e.g.~tables \ref{ener21},~\ref{ener31},~\ref{ener11}, and~\ref{ener51}.
In particular, the improvement becomes better when the total energy of the collision is increased, which 
is expected because the perturbative convergences improves away from the pair-production threshold.

\begin{table}[!h]
 \caption{Top-quark pair production at $\sqrt s = 380$~GeV for unpolarized beams and the polarization configurations of table~\ref{tab:benchpol}.
 The renormalization scale is chosen to be $\mu=\sqrt{s}$, with the scale uncertainties of the symmetric cross sections $\sigma_S$ given by the shifts 
 in the super- and subscripts (corresponding to scales $\mu=2\sqrt{s}$ and $\mu=\sqrt{s}/2$, respectively).
 Symmetric cross sections $\sis$ in units of pb, $\afb$ to LO, and the terms $A_1$, $A_2$ defined in \eqref{afb1ent}, \eqref{afbent} that yield the 
 expanded $\afb$, respectively, to NLO and NNLO QCD. The numbers for $A_1$, $A_2$ and their scale variations are given in the unit of $10^{-2}$.}
 \label{ener21}
\begin{center}
\resizebox{1.0\columnwidth}{!}{
\begin{tabular}{  |c |c c|c c| c c|}
\hline 	  	      												
Beam polarization		&\multicolumn{2}{ c| }{ LO }				&\multicolumn{2}{ c| }{ NLO }							&\multicolumn{2}{ c| }{ NNLO }	\\ \cline{2-7}
$(e^{-}_L, \, e^{+}_R)	$	&$\sigma_S$ $[\rm pb]$		&$\afb^{\rm LO}$		&$\sigma_S$ $[\rm pb]$			&$A_1$ [$10^{-2}$] 								&$\sigma_S$ $[\rm pb]$		&$A_2$ [$10^{-2}$] 	\\ \hline
(0,\, 0)				&0.58477			&0.2342				&$0.78874^{-0.01484}_{+0.01741}$			&$3.67^{+ 0.313}_{-0.267}$				&$0.85037^{-0.01009}_{+0.01002}$			&$2.92^{+0.188}_{-0.168}$		\\  

$(-80\%, \, +30\%)$		&0.32039			&0.2549				&$0.43232^{-0.00814}_{+0.00955}$			&$3.62^{+0.309}_{-0.263}$				&$0.46633^{-0.00556}_{+0.00553}$			&$2.86^{+0.183}_{-0.163}$		\\ 

$(+80\%, \, -30\%)$		&0.56846			&0.2226				&$0.76657^{-0.01441}_{+0.01691}$			&$3.70^{+0.316}_{-0.270}$				&$0.82623^{-0.00977}_{+0.00970}$			&$2.95^{+0.191}_{-0.170}$		\\  

$(+80\%, \,  +30\%)$		&0.99800			&0.2196				&$1.34571^{-0.02530}_{+0.02968}$			&$3.71^{+0.317}_{-0.270}$				&$1.45035^{-0.01714}_{+0.01701}$			&$2.96^{+0.192}_{-0.171}$		\\ 

$(-80\%, \, -30\%)$		&0.45224			&0.2664				&$0.61037^{-0.01151}_{+0.01350}$			&$3.59^{+0.306}_{-0.261}$				&$0.65856^{-0.00788}_{+0.00784}$			&$2.82^{+0.180}_{-0.160}$		\\ 	
\hline	 																											
\end{tabular}}
\end{center} 
\end{table}

\begin{table}[!h]
 \caption{ Top-quark pair production at
 $\sqrt s = 400$~GeV. The meaning of the variables is as in table~\ref{ener21}.
 }
 \label{ener31}
\begin{center}
\resizebox{1.0\columnwidth}{!}{
\begin{tabular}{ | c |c c|c c| c c|}
\hline 	     												
Beam polarization		&\multicolumn{2}{ c| }{ LO }				&\multicolumn{2}{ c| }{ NLO }				&\multicolumn{2}{ c| }{ NNLO }	\\ \cline{2-7}
$(e^{-}_L, \, e^{+}_R)	$	&$\sigma_S$ $[\rm pb]$		&$\afb^{\rm LO}$		&$\sigma_S$ $[\rm pb]$		&$A_1$ [$10^{-2}$] 				&$\sigma_S$ $[\rm pb]$		&$A_2$ [$10^{-2}$] 	\\ \hline
(0,\, 0)				&0.62928			&0.2845				&$0.79311^{-0.01186}_{+0.01389}$			&$3.39^{+0.287}_{-0.245}$				&$0.83400^{-0.00685}_{+0.00648}$			&$2.31^{+0.106}_{-0.101}$		\\  

$(-80\%, \, +30\%)$		&0.34658			&0.3083				&$0.43708^{-0.00655}_{+0.00767}$			&$3.31^{+0.281}_{-0.240}$				&$0.45987^{-0.00381}_{+0.00362}$			&$2.25^{+0.102}_{-0.097}$		\\  

$(+80\%, \, -30\%)$		&0.60992			&0.2710				&$0.76845^{-0.01147}_{+0.01344}$			&$3.43^{+0.291}_{-0.248}$				&$0.80780^{-0.00660}_{+0.00623}$			&$2.35^{+0.109}_{-0.104}$		\\ 

$(+80\%, \,  +30\%)$		&1.06997			&0.2675				&$1.34797^{-0.02012}_{+0.02357}$			&$3.44^{+0.292}_{-0.249}$				&$1.41687^{-0.01156}_{+0.01091}$			&$2.36^{+0.109}_{-0.104}$		\\  

$(-80\%, \, -30\%)$		&0.49064			&0.3215				&$0.61895^{-0.00929}_{+0.01088}$			&$3.27^{+0.277}_{-0.237}$				&$0.65144^{-0.00543}_{+0.00516}$			&$2.21^{+0.099}_{-0.095}$			\\ 	
\hline	 																											
\end{tabular}}
\end{center} 
\end{table}

 \begin{table}[!h]
 \caption{ Top-quark pair production at $\sqrt s = 500$~GeV. The meaning of the variables is as in table~\ref{ener21}.
 }
 \label{ener11} 
\begin{center}
\resizebox{1.0\columnwidth}{!}{
\begin{tabular}{ |c |c c|c c| c c|}
\hline 	  	      												
Beam polarization		&\multicolumn{2}{ c| }{ LO }				&\multicolumn{2}{ c| }{ NLO }				&\multicolumn{2}{ c |}{ NNLO }	\\ \cline{2-7}
$(e^{-}_L, \, e^{+}_R)	$	&$\sigma_S$ $[\rm pb]$		&$\afb^{\rm LO}$		&$\sigma_S$ $[\rm pb]$		&$A_1$ [$10^{-2}$] 				&$\sigma_S$ $[\rm pb]$		&$A_2$ [$10^{-2}$] 	\\ \hline
(0,\, 0)				&0.55084			&0.4169				&$0.62006^{-0.00489}_{+0.00571}$			&$2.26^{+0.186}_{-0.159}$				&$0.63038^{-0.00187}_{+0.00147}$			&$1.16^{+0.004}_{-0.014}$		\\  

$(-80\%, \, +30\%)$		&0.30850			&0.4458				&$0.34762^{-0.00276}_{+0.00323}$			&$2.14^{+0.177}_{-0.151}$				&$0.35362^{-0.00108}_{+0.00086}$			&$1.10^{+0.003}_{-0.013}$		\\ 

$(+80\%, \, -30\%)$		&0.52877			&0.4001				&$0.59488^{ -0.00467}_{+0.00545}$			&$2.32^{+0.192}_{-0.164}$				&$0.60456^{-0.00176}_{+0.00138}$			&$1.20^{+0.004}_{-0.015}$		\\  

$(+80\%, \,  +30\%)$		&0.92534			&0.3957				&$1.04086^{-0.00816}_{+0.00953}$			&$2.34^{+0.193}_{-0.165}$				&$1.05771^{-0.00307}_{+0.00239}$			&$1.21^{+ 0.004}_{-0.015}$		\\ 

$(-80\%, \, -30\%)$		&0.44074			&0.4614				&$0.49689^{-0.00397}_{+0.00463}$			&$2.08^{+0.172}_{-0.147}$				&$0.50564^{-0.00157}_{+0.00126}$			&$1.07^{+0.003}_{-0.012}$		\\ 	
\hline	 																											
\end{tabular}}
\end{center} 
\end{table}
 
 \begin{table}[!h]
 \caption{Top-quark pair production at 
 $\sqrt s = 700$~GeV.  The meaning of the variables is as in table~\ref{ener21}, except that the numbers for $A_2$ and its scale variations are given in the unit of $10^{-3}$.}
 \label{ener51}
 \begin{center}
\resizebox{1.0\columnwidth}{!}{
\begin{tabular}{  |c |c c|c c| c c|}
\hline 	 	      												
Beam polarization				&\multicolumn{2}{ c| }{ LO }				&\multicolumn{2}{ c| }{ NLO }				&\multicolumn{2}{ c| }{ NNLO }	\\ \cline{2-7}
$(e^{-}_L, \, e^{+}_R)$			&$\sigma_S$ $[\rm pb]$		&$\afb^{\rm LO}$		&$\sigma_S$ $[\rm pb]$		&$A_1$ [$10^{-2}$] 				&$\sigma_S$ $[\rm pb]$		&$A_2$ [$10^{-3}$]	\\ \hline
(0,\, 0)						&0.32344			&0.5144				&$0.34560^{-0.00151}_{+0.00176}$			&$0.90^{+0.071}_{-0.061}$				&$0.34759^{-0.00040}_{+0.00024}$			&$4.6^{+0.009}_{-0.048}$		\\  

$(-80\%, \, +30\%)$				&0.18367			&0.5440				&$0.19644^{-0.00087}_{+0.00101}$			&$0.79^{+0.063}_{-0.054}$				&$0.19766^{-0.00024}_{+0.00015}$			&$4.2^{+0.027}_{-0.058}$		\\ 

$(+80\%, \, -30\%)$				&0.30796			&0.4968				&$0.32887^{-0.00143}_{+0.00166}$			&$0.96^{+0.076}_{-0.065}$				&$0.33068^{-0.00037}_{+0.00021}$			&$4.8^{-0.001}_{-0.043}$		\\  
	
$(+80\%, \,  +30\%)$				&0.53779			&0.4922				&$0.57422^{-0.00249}_{+0.00289}$			&$0.97^{+0.077}_{-0.066}$				&$0.57734^{-0.00064}_{+0.00037}$			&$4.9^{-0.004}_{-0.041}$		\\  

$(-80\%, \, -30\%)$				&0.26435			&0.5597				&$0.28288^{-0.00126}_{+0.00147}$			&$0.74^{+ 0.059}_{-0.050}$				&$0.28469^{-0.00035}_{+0.00022}$			&$4.0^{+0.037}_{-0.063}$		\\ 	
\hline	 																											
\end{tabular}}
\end{center} 
\end{table}

\begin{table}[!h]
 \caption{ The factors $C_1$, $C_2$ defined in eq.~(\ref{afb1unex})  and eq.~(\ref{afbgesamt}) that yield the unexpanded $\afb$ to NLO and NNLO QCD. The numbers for $C_1$, $C_2$ and their scale variations are given in the unit of $10^{-2}$.}
 \label{c1c2}
\begin{center}
\resizebox{1.0\columnwidth}{!}{
\begin{tabular}{ |c |c |c|c |c| c |c| }
\hline 	 	      												
\multicolumn{2}{ |c| }{Beam polarization}				&\multirow{2}{*}{(0, ~0)}		&\multirow{2}{*}{$(-80\%,  +30\%)$}	&\multirow{2}{*}{$(+80\%, -30\%)$}	&\multirow{2}{*}{$(+80\%, +30\%)$}		&\multirow{2}{*}{$(-80\%,  -30\%)$}\\ 
\multicolumn{2}{ |c| }{$(e^{-}_L, \, e^{+}_R)$}			&						&								&								&							   &\\  \hline
\multirow{2}{*}{380~GeV}			&$C_1-1$ [$10^{-2}$]			&$2.72^{+0.170}_{-0.149}$		&$2.68^{+0.169}_{-0.148}$		&$2.75^{+0.168}_{-0.152}$		&$2.75^{+0.169}_{-0.152}$			&$2.66^{+0.161}_{-0.145}$\\
							&$C_2-1$	[$10^{-2}$]		&$5.41^{+0.435}_{-0.360}$		&$5.32^{+0.429}_{-0.355}$		&$5.47^{+0.437}_{-0.367}$		&$5.48^{+0.440}_{-0.368}$			&$5.26^{+0.418}_{-0.352}$\\ \hline
 \multirow{2}{*}{400~GeV}		         &$C_1-1$	 [$10^{-2}$]		&$2.69^{+0.180}_{-0.155}$		&$2.63^{+0.177}_{-0.149}$		&$2.72^{+0.179}_{-0.160}$		&$2.73^{+0.180}_{-0.158}$			&$2.59^{+0.174}_{-0.151}$\\
							&$C_2-1$ [$10^{-2}$]			&$4.96^{+0.374}_{-0.316}$		&$4.84^{+0.364}_{-0.305}$		&$5.03^{+0.379}_{-0.323}$		&$5.05^{+0.381}_{-0.322}$			&$4.77^{+0.356}_{-0.304}$\\ \hline
\multirow{2}{*}{500~GeV}			&$C_1-1$	[$10^{-2}$]		&$2.01^{+0.146}_{-0.129}$		&$1.90^{+0.136}_{-0.121}$		&$2.06^{+0.152}_{-0.133}$		&$2.08^{+0.152}_{-0.133}$			&$1.85^{+0.136}_{-0.116}$	\\
							&$C_2-1$	[$10^{-2}$]		&$3.23^{+0.200}_{-0.178}$		&$3.07^{+0.193}_{-0.168}$		&$3.33^{+0.206}_{-0.183}$		&$3.36^{+0.208}_{-0.183}$			&$2.98^{+0.184}_{-0.161}$\\	\hline						
\multirow{2}{*}{700~GeV}			&$C_1-1$	[$10^{-2}$]		&$0.84^{+0.060}_{-0.055}$		&$0.74^{+0.053}_{-0.050}$		&$0.90^{+0.066}_{-0.057}$		&$0.91^{+0.065}_{-0.058}$			&$0.69^{+0.051}_{-0.044}$\\
							&$C_2-1$	[$10^{-2}$]		&$1.32^{+0.075}_{-0.068}$		&$1.17^{+0.072}_{-0.063}$		&$1.40^{+0.082}_{-0.069}$		&$1.42^{+0.079}_{-0.071}$			&$1.10^{+0.066}_{-0.058}$\\	 \hline																											
\end{tabular}}
\end{center} 
\end{table}
  
Moreover, for fixed c.m. energy, the QCD correction terms to $\afb^{\rm LO}$ show a marginal dependence on the beam polarization.
In order to quantify the variations of the expansion terms $A_1,\, A_2$ with the beam polarization, we introduce the ratio%
\begin{equation} \label{eq:Ri}
R_i (e^-_{L}, e^+_{R}) = \frac{A_i(e^-_{L}, e^+_{R}) - A_i(0, 0)}{A_i(0, 0)} \, ,
\end{equation}%
where $A_i(e^-_{L}, e^+_{R})$ $(i=1,2)$ denote the $i$-th order QCD correction terms for the electron and positron polarization configurations,
specified  by $e^-_{L}$ and $e^+_{R}$, respectively, as given in table~\ref{tab:benchpol}, and in particular $A_i(0, 0)$ are the QCD correction terms for unpolarized beams. 
Furthermore, we use
\begin{equation}\label{eq:rimax}
R_i^{\mathrm{max}} \equiv \mathrm{max}[R_i (e^-_{L}, e^+_{R})] - \mathrm{min}[R_i (e^-_{L}, e^+_{R})] 
 = \frac{\mathrm{max}[A_i (e^-_{L}, e^+_{R})] - \mathrm{min}[A_i (e^-_{L}, e^+_{R})]}
{A_i(0, 0)}
\end{equation}%
for signifying the maximal spread of $A_i$ in relation to the correction for unpolarized beams. 
The meaning of the max/min operation is as usual, for example, $\mathrm{max/min}[A_i (e^-_{L}, e^+_{R})]$ denotes the maximal/minimal value of the QCD correction term $A_i(e^-_{L}, e^+_{R})$ 
for all beam polarization configurations $e^-_{L}, e^+_{R}$ considered in this paper. The same remark applies to the notation $\mathrm{max/min}[R_i (e^-_{L}, e^+_{R})]$.
The values of $R_i^{\mathrm{max}}$ are given in table~\ref{table:Rmax} for the four benchmark polarizations and the c.m. energies considered.
For instance, for  $\sqrt{s}=500$~GeV $A_1, A_2$ become maximal for the polarization configuration $e^-_{L} = 80\%,\, e^+_{R} = 30\%$ where $R_1(80\%,\, 30\%) =4.63\%$, $R_2(80\%,\, 30\%) =5.17\%$, and the minimal values are taken  at $e^-_{L} = -80\%,\, e^+_{R} = -30\%$ where $R_1(80\%,\, 30\%) = -7.7\%$, $R_2(80\%,\, 30\%) = -8.0\%$. 
Although the relative spreads listed in table~\ref{table:Rmax} appear to be quite large, one should notice that in absolute terms they amount to changes of the NLO and NNLO QCD correction factors for unpolarized beams, $(1+A_1)$ and $(1+A_1 + A_2)$, respectively, of only a few per mille.
The projected statistical uncertainty of the polarized $\tbart$ cross section and top forward-backward asymmetry at future high-energy high-luminosity electron-positron 
colliders was estimated to be at the level of a few per mille, c.f.~\cite{ILDConceptGroup:2020sfq,CLICdp:2018esa}, hence the effect of the beam polarization on the QCD correction factors are comparable to that.
With the results for the symmetric $\tbart$ cross section and the forward-backward asymmetry with polarized beams presented in tables~\ref{ener21},~\ref{ener31},~\ref{ener11}, 
the corresponding theoretical uncertainties can be reduced.

 \begin{table}[!h]
 \caption{The ratio $R_i^{\mathrm{max}}$ for different c.m. energies.}
 \label{table:Rmax}
\begin{center}
\begin{tabular*}{0.60 \textwidth}{|@{\extracolsep{\fill} } c| c| c| c| c| }
\hline  
   $\sqrt{s}$~[GeV]     &380     & 400   & 500     & 700     \\ \hline   
  $R_1^{\mathrm{max}}$ &3.4\%   &5.0\%  &12.3\%    &26.0\%    \\  
  $R_2^{\mathrm{max}}$ &4.6\%   &6.2\%  &13.3\%    &20.3\%    \\ \hline  
\end{tabular*}
\end{center}
\end{table}

The dependence on the beam polarization arises because the final-state quark is massive and because of the coherent contributions of the $Z$- and $\gamma$-exchange 
contributions. Specifically, as far as $\afb$ in its expanded form is concerned, the beam-polarization dependence of the $A_i$ 
originate from the ratios of the symmetric cross sections (that depend both on $P_v$ and $P_a$) in eqs.~\eqref{eq:afb1},~\eqref{eq:afb2}.
The dependence of the antisymmetric cross sections on the beam polarization is just an overall factor that cancels in the respective ratios in
eqs.~\eqref{eq:afb1},~\eqref{eq:afb2}.
In addition, starting at NNLO QCD, there are also flavor singlet contributions that supply additional polarization dependence.
However, for $\tbart$ production at energies considered here the singlet contributions are tiny (of order $10^{-4}$ relative to $\afb^{\rm LO}$).
In other words,  $\tbart$ production at these energies is completely dominated by the flavor non-singlet corrections. 
In the limit of extremely high energies, $m_t/\sqrt{s} \to 0$, the polarization dependence of the $A_i$ from the non-singlet contribution disappears.

\begin{figure}[!htp]
\begin{center}
\subfigure[]{\label{fig1a}
\includegraphics[width=0.45\textwidth]{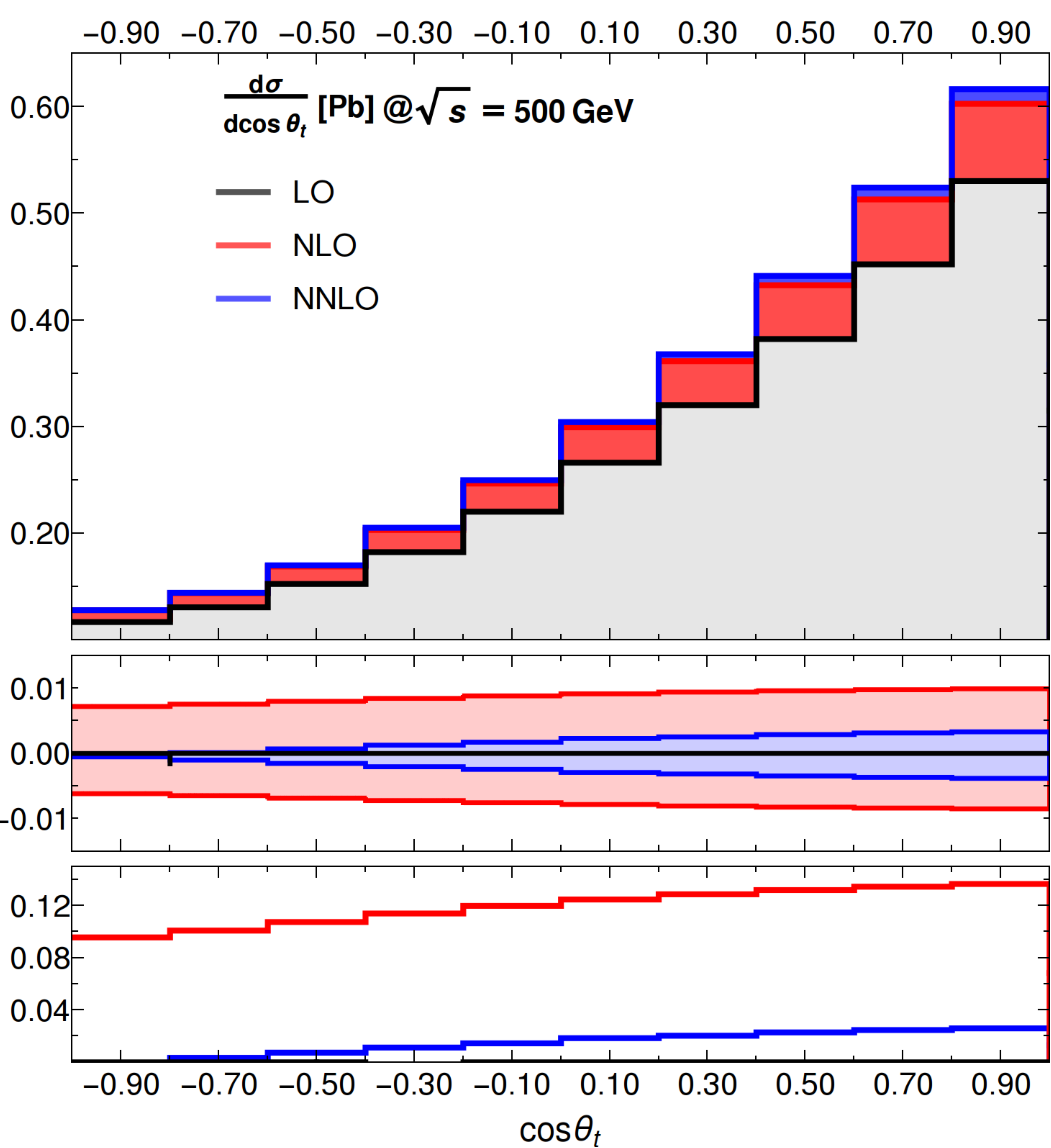} }
\subfigure[]{\label{fig1b}
\includegraphics[width=0.45\textwidth]{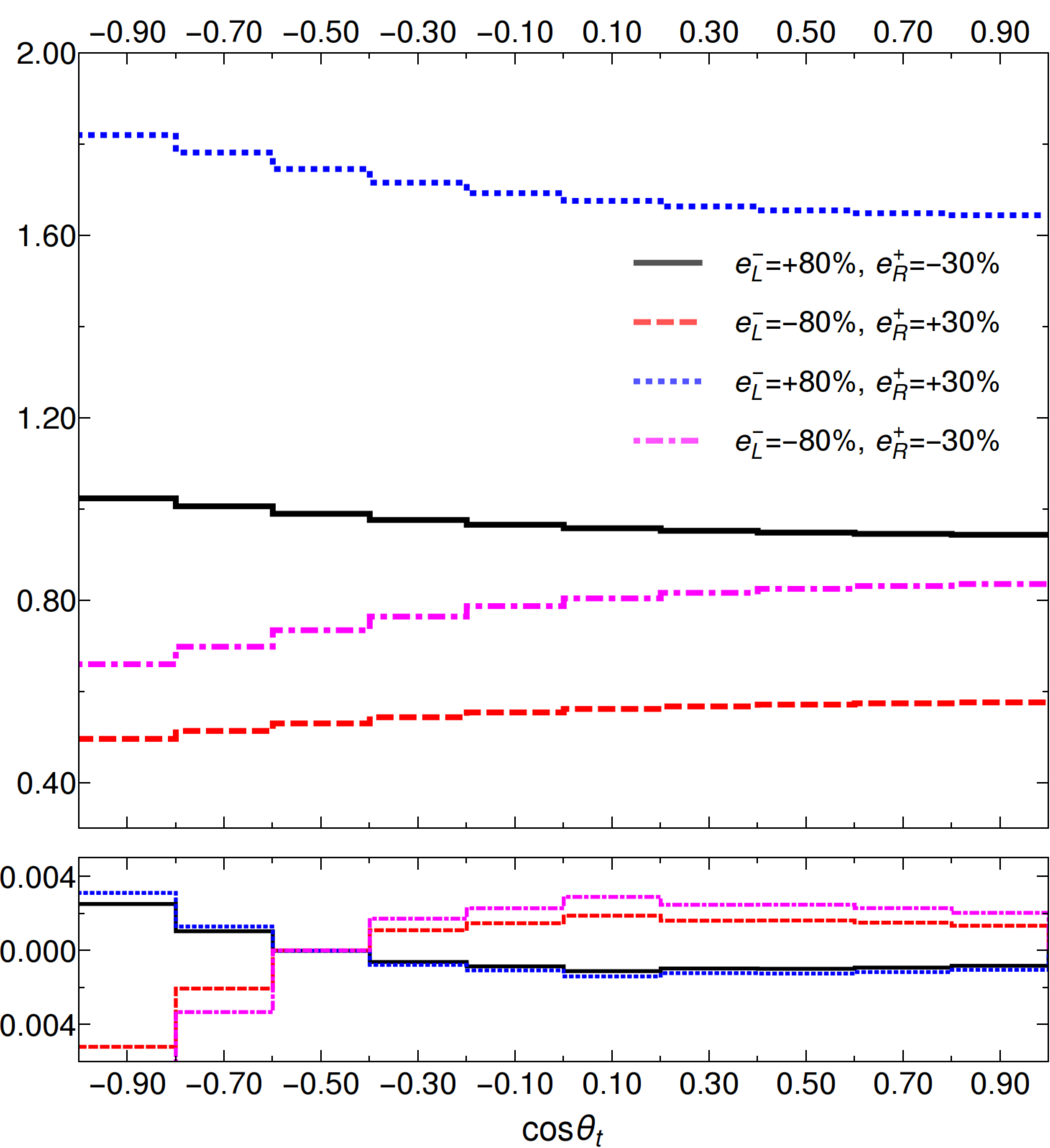} }
\caption{(a) The upper panel shows the  $\cos \theta_t$ distribution for unpolarized beams at LO(grey, lower steps), NLO(red, steps in the middle), and NNLO QCD(blue, upper steps) for $\mu = \sqrt{s} = 500$~GeV. 
The panel in the middle displays the scale variations $[d\sigma_{\rm NLO}(\mu^{'})/d\sigma_{\rm NLO}(\mu=\sqrt{s})]-1$ (red, wide band) and  
$[d\sigma_{\rm NNLO}(\mu^{'})/d\sigma_{\rm NNLO}(\mu=\sqrt{s})]-1$ (blue, narrow band) of the first and second order QCD corrections, where $\sqrt s/2 < \mu^{'} < 2\sqrt s$. 
The lower panel shows the ratio $d\sigma_1/d\sigma_{\rm LO}$ (red, upper steps) and $d\sigma_2/d\sigma_{\rm LO}$ (blue, lower steps) for $\mu = \sqrt s$. 
(b) The upper panel shows the LO QCD ratios of the $\cos \theta_t$ distributions with polarized and unpolarized beams  at $\sqrt{s}=500$~GeV. 
 The lower panels display the ratios of the $\cos\theta_t$ distributions
$[(d\sigma_{\rm NNLO}^{\rm pol}/d\sigma_{\rm LO}^{\rm pol})/(d\sigma_{\rm NNLO}/d\sigma_{\rm LO})] -1$ for the various polarization configurations for $\mu=\sqrt{s}$.
The  coding is the same as in the upper panel.
}
\label{fig1}
\end{center}
\end{figure}

We investigated also the dependence of $A_1, A_2$ on the uncertainty of the value of top-quark pole mass. Varying $\delta m_t = \pm 0.7$~GeV we found that the 
resulting variation of these correction terms given in tables~\ref{ener21},~\ref{ener31},~\ref{ener11}, and~\ref{ener51} is less than $1\%$.

The top-quark $\afb$ was computed before in \cite{Gao:2014eea,Chen:2016zbz} to NNLO QCD  for unpolarized beams at $\sqrt{s}=500$~GeV.
Using the same values of the input parameters that were used in these papers we checked that we agree with their results.

The above forward-backward asymmetries are obtained from the respective distribution of the top-quark polar angle $\theta_t =\angle(\km,\qm)$ in the $e^+e^-$
 c.m. frame, cf. eq.~\eqref{eeQQX}.
Here we restrict ourselves to discuss its distribution for $\tbart$ production at 500~GeV.
In the plots below we use the notation $d\sigma_{\rm NLO} = d\sigma_{\rm LO} + d\sigma_1$ and  $d\sigma_{\rm NNLO} = d\sigma_{\rm LO} + d\sigma_1 + d\sigma_2$ 
for the differential cross section at NLO and NNLO QCD, respectively.
The upper panel of the plot in figure~\ref{fig1}a displays the distribution of $\cos\theta_t$  at LO,
NLO, and NNLO QCD for unpolarized beams.  The panel in the middle of the plot shows that the
inclusion of the order $\as^2$ correction significantly reduces the dependence of this distribution
on variations of the scale $\mu$. 
The lower panel exhibits the ratio  $d\sigma_1/d\sigma_{\rm LO}$  and $d\sigma_2/d\sigma_{\rm LO}$  for $\mu = \sqrt s$. 
Both the order $\as$ and order $\as^2$ corrections 
 follow the same pattern as the leading-order distribution: they are larger in the top-quark forward direction and thus increase the forward-backward asymmetry.  
The upper panel of the plot in figure~\ref{fig1}b displays, at $\sqrt{s}=500$~GeV and LO QCD, the ratios of the $\cos \theta_t$ distributions with and without polarized beams.
The lower panel shows the ratios of the $\cos\theta_t$ distributions
$[(d\sigma_{\rm NNLO}^{\rm pol}/d\sigma_{\rm LO}^{\rm pol})/(d\sigma_{\rm NNLO}/d\sigma_{\rm LO})] -1$ for the various polarization configurations for $\mu=\sqrt{s}$.

 Finally we consider the polarized forward-backward asymmetry  $\afblr$  defined in \eqref{defalrfb}.  It is independent of the polarization configurations   of table~\ref{tab:benchpol}. 
 Moreover, for all c.m. energies, the QCD correction terms $A^{\rm{pol,FB}}_1$ and $A_2^{\rm{pol,FB}}$ to the leading-order
 polarized forward-backward asymmetry are the same as the respective unpolarized terms $A_1, A_2$. The respective ratios $A_{\rm FB}/\afblr$ are given in
 table~\ref{ratPolUn}. They vary slightly with the c.m. energy due to the slight change in the relative weights of the coherent $Z$-boson and photon
  exchange amplitudes.

\begin{table}[!h]
 \caption{The ratio of the polarized forward-backward asymmetry and the asymmetry for unpolarized beams for top-quark pair production at  various
  c.m. energies. The respective ratio at $\sqrt{s}$ holds for all  polarization configurations of table~\ref{tab:benchpol} and is valid at LO, NLO, and NNLO QCD.}
 \label{ratPolUn}
\begin{center}
\begin{tabular*}{0.60 \textwidth}{|@{\extracolsep{\fill} } c| c| c| c| c| }
\hline 
  $\sqrt{s}$~[GeV]  	     & 380			& 400			&  500	 		& 700	   \\ \hline	
  $A_{\rm FB}/\afblr$ & 3.053			& 3.065			 & 3.103			& 3.136 \\ \hline	 
\end{tabular*}
\end{center} 
\end{table}

\section{Asymmetries for $b$-quark pair production at the Z peak}
\label{sec:asyb}

Next we consider the the production of $\bbarb$ pairs at the $Z$ resonance to NNLO QCD and to lowest order in the electroweak couplings, i.e., $\bbarb$ production by a virtual 
photon is not taken into account. The $b$ quark is taken to be massive, with an on-shell mass value given in \eqref{eq:parameter} where also the other SM parameter values 
relevant for the calculations below are listed.

As in the case of top-quark production we consider  $\bbarb$ production both by unpolarized and polarized $e^+$ and $e^-$ beams, with  polarization configurations as given in table~\ref{tab:benchpol}.

We will compute the $b$-quark $\afb$ for several definitions of the  forward and backward hemispheres.
First we will use the  $b$-quark direction of flight and the oriented thrust axis
  for defining these hemispheres.
 If the $b$-quark direction of flight is chosen  then  the hemispheres are separated according to $\cos\theta_b$ being larger or smaller than zero, where 
 $\theta_b =\angle(\km,\qm)$, cf. eq.~\eqref{eeQQX}.
 Because an accurate determination of the $b$-quark flight direction is difficult,
   experimental analyses in the past often used
   the thrust axis as  reference axis. For a 
 given $n$-parton event described by a collection of final-state
 four-momenta $\{k_i\}_{i=1}^{n}$ (related by momentum conservation), 
 the thrust axis is the direction $\nt$  that
 maximizes the thrust $T$ defined by \cite{Farhi:1977sg,Brandt:1964sa,Brandt:1978zm}:
\begin{equation}\label{eqThrustdef}
T=\max\limits_{\nt}\frac{\sum\limits_{i=1}^n|\ki\cdot\nt|}{\sum\limits_{i=1}^n|\ki|},\qquad |\nt|=1.
\end{equation}
 The orientation of the thrust axis is fixed by requiring $\nt\cdot\km>0$. 
  If the thrust axis is chosen as  reference axis, 
 the forward and backward hemispheres are discriminated by the sign of $\cos\theta_T$ where 
  $\theta_T=\angle(\nt, \qm)$.    

We recall that the various contributions to the inclusive $b$-quark cross section and, in particular, to the $b$-quark FB asymmetry at NNLO QCD can be classified into 
flavor non-singlet, flavor singlet, and interference or  triangle terms. (For details see, e.g., \cite{Bernreuther:2016ccf}.) The contribution from the $\bbarb\bbarb$
 final state deserves special mention. In the calculation of the $b$-quark FB asymmetry, i.e., in the calculation of the  forward and backward cross sections  $\sis$ and $\sia$,
 we have the following situations:
 i) both $b$ quarks are in the forward (backward) direction, thus they  contribute both to $\sigma_F$ $(\sigma_B$).
ii) one $b$ quark is forward, the other one backward, thus there is a contribution both to $\sigma_F$ and $\sigma_B.$

Table \ref{table:quark_axis} shows, for the $b$-quark  axis definition of the forward and backward hemispheres, our results 
 for unpolarized beams and for the polarization configurations of table~\ref{tab:benchpol}. 
The cross sections $\sis$ at LO, NLO, and NNLO QCD are given for the  scale choice $\mu = m_Z$. Apart from the Born level  $\afb^{\rm{LO}}$ also the
 first- and second-order expansion terms are displayed, including the changes that result from varying $\mu$ between $m_Z/2$ and $2 m_Z$. For brevity we display here and in the tables below
  only the QCD corrections
  to $\afb^{\rm{LO}}$  in expanded form. 
  
  \begin{table}[!h]
 \caption{ 
 Production of $\bbarb$ at the $Z$ peak  for unpolarized beams and the polarization configurations of table~\ref{tab:benchpol}. Here the FB asymmetry is defined with 
 respect to the $b$-quark axis. 
 Symmetric  cross sections  $\sis$ in units of pb,  $\afb$ to LO and the terms $A_1$, $A_2$ defined in \eqref{afb1ent}, \eqref{afbent} that yield 
  the expanded $\afb$ to NLO and NNLO QCD. The parameters listed in eq.~\eqref{eq:parameter} are used.
 The renormalization scale is chosen to be $\mu= m_Z$.
 The numbers in superscript (subscript) refer to the changes if $\mu = 2 m_Z$ ($\mu = m_Z/2$) is chosen. The numbers for $A_1$, $A_2$ and their scale variations are given in the unit of $10^{-2}$.}
 \label{table:quark_axis}
 \begin{center}
\resizebox{1.0\columnwidth}{!}{
\begin{tabular}{  |c |c c|c c| c c|}
\hline 	 	      												
Beam polarization		&\multicolumn{2}{ c| }{ LO }				&\multicolumn{2}{ c| }{ NLO }						&\multicolumn{2}{ c| }{ NNLO }	\\ \cline{2-7}
$e^{-}_L, \,e^{+}_R$		&$\sigma_S$ $[\rm pb]$		&$\afb^{\rm LO}$		&$\sigma_S$ $[\rm pb]$		&$A_1$ [$10^{-2}$]		&$\sigma_S$ $[\rm pb]$		&$A_2$ [$10^{-2}$]	\\ \hline
(0, \, 0)				&8747.4			&0.1512				&$9122.8^{-35.5}_{+44.1}$			&$-2.92^{+0.277}_{-0.343}$	&$9164.8^{-12.4}_{+7.7}$			&$-1.28^{+0.046}_{-0.008}$			\\  

($-80\%, \,+30\%$)		&5710.7			&-0.3644				&$5955.8^{-23.2}_{+28.8}$			&$-2.92^{+0.277}_{-0.343}$	&$5990.6^{-9.4}_{+6.9}$			&$-1.35^{+0.028}_{+0.005}$			\\  

($+80\%, \,-30\%$)		&7585.3			&0.5394				&$7910.9^{-30.8}_{+38.2}$			&$-2.92^{+0.277}_{-0.343}$	&$7939.9^{-9.4}_{+4.8}$			&$-1.16^{+0.077}_{ -0.031}$			\\ 

($+80\%, \,+30\%$)		&12908.8			&0.6531				&$13462.8^{-52.4}_{+65.0}$			&$-2.92^{+0.277}_{-0.343}$	&$13508.5^{-15.3}_{+7.3}$			&$-1.12^{+0.085}_{-0.036}$			\\  

($-80\%, \,-30\%$)		&8784.7			&-0.5862				&$9161.7^{-35.7}_{+44.2}$			&$-2.92^{+0.277}_{-0.343}$	&$9220.1^{-15.3}_{+11.8}$			&$-1.42^{+0.016}_{+0.013}$			\\ 	
\hline	 																											
\end{tabular}}
\end{center} 
\end{table}

 As the numbers in this table show, $A_1$ is, in contrast to the case of the top quark, independent of the beam polarization. This is because  only the $Z$-boson exchange 
 is taken into account. The polarization dependence of $A_2$ arises solely from the flavor singlet contributions. Defining a quantity
  $R_2^{\mathrm{max}}(b)$ in analogy to eq.~\eqref{eq:rimax} in order to quantify the maximal spread of $A_2$ due to beam polarization one gets 
  $|R_2^{\mathrm{max}}(b)| = 23\%$. As already found  in \cite{Bernreuther:2016ccf} for unpolarized beams, the order $\as^2$ corrections are quite large for all polarization 
  configurations. From table~\ref{table:quark_axis}  we get $40\% \leq A_2/A_1 \leq 48\%$.
  
The flavor non-singlet corrections $A_2^{\rm ns}$ of order $\alpha_s^2$ that contain the neutral current couplings of the $b$ quarks, are
 independent of beam  polarization.
For the quark axis definition of the asymmetry  we get $A_2^{\rm ns}=-0.0084$ (for $\mu=m_Z$) that amounts to $65\%$ of the total correction $A_2$.

  Table~\ref{table:thrust_axis} contains the analogous information for the definition of the forward and backward hemispheres with respect to the oriented thrust axis. 
  Using again eq.~\eqref{eq:rimax} to quantify the maximal spread of $A_2$ for the above beam polarizations, we get here $|R_2^{\mathrm{max}}(b)| = 27\%$.
  The ratio $A_2/A_1$ varies between $33\%$ and $43\%$ which is somewhat smaller than in the case of the $b$-quark axis definition. 

\begin{table}[!htbp]
 \caption{ Same as table~\ref{table:quark_axis}, but here the asymmetry is defined with respect to  the thrust axis.}
 \label{table:thrust_axis}
 \begin{center}
 \resizebox{1.0\columnwidth}{!}{
\begin{tabular}{  |c |c c|c c |c c|}
\hline 	 	      												
Beam polarization		&\multicolumn{2}{ c| }{ LO }				&\multicolumn{2}{ c| }{ NLO }						&\multicolumn{2}{ c| }{ NNLO }	\\ \cline{2-7}
$e^{-}_L, \,e^{+}_R$		&$\sigma_S$ $[\rm pb]$	&$\afb^{\rm LO}$	&$\sigma_S$ $[\rm pb]$		&$A_1$  [$10^{-2}$]		&$\sigma_S$ $[\rm pb]$		&$A_2$  [$10^{-2}$]	\\ \hline
(0, \, 0)				&8747.4			&0.1512				&9122.8			&$-2.88^{+0.273}_{-0.338}$	&9164.8			&$-1.11^{+0.085}_{ -0.037}$			\\  

$(-80\%, \,+30\%)$		&5710.7			&-0.3644				&5955.8			&$-2.88^{+0.273}_{-0.338}$	&5990.6			&$-1.19^{+0.064}_{-0.022}$			\\  

($+80\%, \,-30\%$)		&7585.3			&0.5394				&7910.9			&$-2.88^{+0.273}_{-0.338}$	&7939.9			&$-0.99^{+0.115}_{-0.058}$			\\ 

($+80\%, \,+30\%$)		&12908.8			&0.6531				&13462.8			&$-2.88^{+0.273}_{-0.338}$	&13508.5			&$-0.96^{+0.122}_{-0.064}$			\\  

($-80\%, \,-30\%$)		&8784.7			&-0.5862				&9161.7			&$-2.88^{+0.273}_{-0.338}$	&9220.1			&$-1.25^{+0.049}_{ -0.338}$			\\ 	
\hline	 																											
\end{tabular}}
\end{center} 
\end{table}%

The $\bbarb\bbarb$ final state has a distinctive experimental signature. Depending on the $b$-tagging efficiency it could, in principle, 
be separated from the other final states. In order to assess the contribution of the $\bbarb\bbarb$ final state, we compute now $\afb$ without including these events. This changes 
   $\sis$ at NNLO QCD and the expansion term $A_2$. The resulting values for these quantities are given in table~\ref{table:withoutQQQQ}
    for unpolarized beams.
 Comparing with the numbers in tables~\ref{table:quark_axis} and~\ref{table:thrust_axis} shows that  $A_2$ is reduced in magnitude by $10\%$ and $11\%$
 in the case of the $b$-quark axis and thrust axis definition, respectively.
 
 \begin{table}[!h]
 \caption{ The $\bbarb$ cross sections $\sis$ in units of pb at NNLO QCD at the $Z$ peak for unpolarized beams and $\mu=m_Z$ 
 without the contribution from the $\bbarb\bbarb$ final state 
 and the resulting expansion terms $A_2$ for the quark-axis and thrust-axis definition of the FB asymmetry. The numbers for $A_1$, $A_2$ and their scale variations are given in the unit of $10^{-2}$.
 }
 \label{table:withoutQQQQ}
\begin{center}
\begin{tabular*}{0.6 \textwidth}{|@{\extracolsep{\fill} } c c| c c|  }
\hline  
                             				 	 \multicolumn{2}{ |c| }{ quark axis }  			&\multicolumn{2}{ c| }{ thrust axis } 	\\ \hline
								 $\sigma_S$ $[\rm pb]$ 		&  $A_2$ [$10^{-2}$]		 & $\sigma_S$ $[\rm pb]$  	&  $A_2$ [$10^{-2}$]     	\\ \hline			
                        				      		9148.0			&$-1.17^{+0.074}_{-0.028}$				&9148.0			&$-0.99^{+0.113}_{-0.057}$
 \\ \hline 
\end{tabular*}
\end{center} 
\end{table}

The forward-backward asymmetry at the $Z$ resonance were computed in  \cite{Bernreuther:2016ccf} with respect to
 the $b$-quark direction and the oriented thrust direction, for massive $b$ quarks and unpolarized beams, with input parameters that differ somewhat
from the ones used here. We agree with these results.

Next we consider jets and use the direction of the $b$ jet to define the forward and backward hemispheres and compute $\afb$. We use the Durham \cite{Catani:1991hj}
and the flavor-$k_T$ \cite{Banfi:2006hf} jet clustering algorithms. 
In applications to flavored massless quark jets, the Durham algorithm is infrared-unsafe  at order $\as^2$  while the flavor-$k_T$ algorithm 
is infrared safe
by construction \cite{Banfi:2006hf}. However, as we consider massive $b$ quarks, also the Durham algorithm allows for an infrared-safe definition of a $b$ jet.
One assigns the flavor number $+1$ $(-1)$ to a $b$ quark ($\bar{b}$ quark) and flavor number $0$ to the other quarks and the gluon. Flavor numbers are added. If two $b$ quarks ($b$ and
  $\bar{b}$) are combined the resulting pseudoparticle has flavor number $2$ $(0)$.
We recall the respective distance measure between every pair of partons (resp. pseudoparticles) $i, j$:
\begin{equation}
 y_{ij}^X = (1-\cos\theta_{ij}) \frac{2 r_X}{s} \, ,
 \label{eq:dismeas}
\end{equation}
where $\theta_{ij}$ is the angle between (pseudo)particles $i$ and $j$. The Durham algorithm is defined by $r_D = {\rm min}(E_i^2,E_j^2)$, where $E_i$ is the energy 
of (pseudo)particle $i$,
while in the case of the  flavor-$k_T$ algorithm 
\begin{equation}
  \label{eq:flavkt}
  r_F = \left\{
    \begin{array}[c]{ll}
      \max(E_i^2, E_j^2)\,, & \quad\mbox{if softer of $i,j$ is flavored,}\\
      \min(E_i^2, E_j^2)\,, & \quad\mbox{if softer of $i,j$ is flavorless.}
    \end{array}
  \right.
\end{equation}
For recombining $i$ and $j$ whose distance  $y_{ij}^X$ is smaller than a specified jet resolution parameter $y_{cut}$ we use the $E$ scheme
 that sums the four-momenta  $(k_{(ij)}=k_i + k_j)$.
  
 Table~\ref{table:bjet_axis}  contains, for unpolarized beams, our results for the  bottom quark cross section $\sis$  to order $\as^2$ where the forward 
  and backward hemispheres are defined
  with respect to the $b$-jet direction, both for the flavor-$k_T$ and the Durham algorithm with a sequence of  jet resolution parameters $y_{cut}$. 
  Moreover, the NLO and NNLO QCD  correction terms to $\afb^{\rm LO}$ are given. The term $A_1$ $(A_2)$ receives contributions from two- and three-jet (two-, three-, and four-jet)
  events with $b$-flavor number larger than zero. Jet events with $b$-flavor number zero are not taken into account; in particular, they are not included in $\sis$. That is why the numbers for 
  $\sis$ differ for different jet algorithms.

  With less stringent jet resolution parameter $y_{cut}$ the magnitudes of the QCD correction factors to the inclusive $b$-jet $A_{FB}$ become smaller. 
  The order $\as^2$ correction terms $A_2$ decrease relative to $A_1$ with increasing $y_{cut}$ as the ratios $A_2/A_1$ listed in table~\ref{table:Ratio} show.
 For  $y_{cut}\geq 0.1$ these ratios become smaller compared to $A_2/A_1$ in the  case of $b$-quark and thrust axis definition of $\afb$, cf. Tables~\ref{table:quark_axis},~\ref{table:thrust_axis}.
 
  From the numbers in table~\ref{table:bjet_axis_no4Q} we deduce that using the Durham (flavor-$k_T$) algorithm the contribution 
   of the $\bbarb\bbarb$ final state to $A_2$ is about $6\%$ ($8\%$) for $y_{cut}=0.01$ and decreases to about $3\%$ ($6\%$) for  $y_{cut}=0.15$.
  
    \begin{table}[!h]
 \caption{ 
 Production of $\bbarb$ at the $Z$ peak  for unpolarized beams. Here the FB asymmetry is defined with 
 respect to the $b$-jet axis and two different jet algorithms are used. 
 Symmetric  cross sections  $\sis$ in units of pb,  $\afb$ to LO and the terms $A_1$, $A_2$ defined in \eqref{afb1ent}, \eqref{afbent} that yield 
  the expanded $\afb$ to NLO and NNLO QCD. The parameters listed in eq.~\eqref{eq:parameter} are used.
 The renormalization scale is chosen to be $\mu= m_Z$. The numbers for $A_1$, $A_2$ and their scale variations are given in the unit of $10^{-2}$.}
 \label{table:bjet_axis}
\begin{center}
\begin{tabular}{  |c |c c c c  c c|}
\hline 	  	     
Jet algorithms						&\multicolumn{2}{ c }{ LO }				&\multicolumn{2}{ c }{ NLO }						&\multicolumn{2}{ c| }{ NNLO }	\\ \cline{2-7}
($y_{cut} $)						&$\sigma_S$ $[\rm pb]$		&$\afb^{\rm LO}$	&$\sigma_S$ $[\rm pb]$		&$A_1$ [$10^{-2}$]		&$\sigma_S$ $[\rm pb]$		&$A_2$ [$10^{-2}$]	\\  \hline
 									
Flavor $k_T$, 0.01		 			&8747.4			&0.1512				&9120.7 			&$-2.88^{+0.037}_{+0.021}$						& 9150.4       		&$-1.09^{+0.216}_{-0.303}$	 \\

Flavor $k_T$, 0.05 					&8747.4			&0.1512				&9111.4 			&$-2.67^{+0.034}_{+0.020}$						& 9121.4			&$-0.83^{+0.167}_{-0.236}$	 \\  																									                     									 
			   								
Flavor $k_T$, 0.10					&8747.4			&0.1512				& 9098.2			&$-2.42^{+0.031}_{+0.018}$					&9097.7			&$-0.68^{+0.138}_{-0.195}$	 \\ 	

Flavor $k_T$, 0.15					&8747.4			&0.1512				&9082.8			&$-2.18^{+0.028}_{+0.016}$					&9075.3 			&$-0.59^{+0.120}_{-0.169}$	\\      												
\hline
Durham, 0.01						&8747.4			&0.1512				&9100.2			&$-2.58^{+0.033}_{+0.019}$						&9112.1 			&$-0.87^{+0.174}_{-0.244}$	\\  	
		
Durham, 0.05 						&8747.4			&0.1512				&9050.4			&$-1.84^{+0.024}_{+0.013}$						&9035.8 			&$-0.52^{+0.105}_{-0.147}$  \\   	

Durham, 0.10						&8747.4			&0.1512				&9018.2			&$-1.46^{+0.019}_{+0.011}$						&8992.4 			&$-0.38^{+0.076}_{-0.107}$  \\   	
															
Durham, 0.15						&8747.4			&0.1512				&8996.7			&$-1.26^{+0.016}_{+0.009}$						& 8964.4			&$-0.33^{+0.065}_{-0.091}$  \\     \hline  
\end{tabular}
\end{center} 
\end{table}%

\begin{table}[!h]
\caption{The ratio $A_2/A_1$ for the two jet algorithms and the jet resolution parameters $y_{cut}$ from table~\ref{table:bjet_axis}. }
\label{table:Ratio}
\begin{center}
\begin{tabular}{  |c   |c  |c  |c  |c| }
\hline	 	   
Flavor~$k_T, y_{cut}$		&0.01		&0.05		&0.10		&0.15	\\ 												
$A_2/A_1$				&37.9\%		&31.1\%		&28.1\%		&27.1\%			  \\ \hline
$\rm{Durham}$, $y_{cut}$		&0.01	&0.05		&0.10		&0.15	\\ 												
$A_2/A_1$			& 33.8\%	&28.5\%		&26.2\%		&26.0\%		  \\ 	\hline	 																					
\end{tabular}
\end{center} 
\end{table}

\begin{table}[!h]
 \caption{ 
 The cross sections $\sis$  in units of pb at NNLO QCD at the $Z$ peak for unpolarized beams and $\mu=m_Z$, with the forward and backward hemispheres
  defined with respect to the $b$-jet axis,
  without the contribution from the $\bbarb\bbarb$ final state 
 and the resulting expansion terms $A_2$  of the FB asymmetry for two jet algorithms. }
 \label{table:bjet_axis_no4Q}
\begin{center}
\begin{tabular}{  |c| c c   c c|   }
\hline  
\multicolumn{1}{ |c }{\multirow{2}{*}{~~$y_{cut}$~~}}        		 	  & 	\multicolumn{2}{ |c }{ Flavor $k_T$ } 				& \multicolumn{2}{ c| }{Durham} 	  	\\ \cline{2-5}
				   		 &$\sigma_S$ $[\rm pb]$		&$A_2$				&$\sigma_S$ $[\rm pb]$		&$A_2$			 \\ \hline
		0.01 	 			   &9136.3				&-0.0101					&9101.4			&-0.0082	\\ 
	    	
		0.05  			    &9111.1			&-0.0079					&9027.9			& -0.0051	\\ 
						
		0.10 	 			   &9088.7				&-0.0065					&8985.1			& -0.0037 	\\ 
				
		0.15 	 			   &9066.7				&-0.0057					&8957.3			&-0.0032 \\ \hline 	 				
\end{tabular}
\end{center} 										
\end{table}

  Next we select two-jet events with the flavor-$k_T$ and Durham algorithm and a specified resolution parameter $y_{cut}$. As in the inclusive case just discussed the axis of the 
  $b$ jet defines whether a jet lies in the forward or backward hemisphere. We calculate the resulting two-jet forward-backward asymmetry to NNLO QCD.
   The tree level-values $\afb^{\rm LO}$ are of course the same as in table~\ref{table:bjet_axis}, and the NLO and NNLO QCD correction terms $A_1^{2j}$, $A_2^{2j}$
    are given in table~\ref{table:bjet_AswithQQQQ}. Comparing with the respective numbers in table~\ref{table:bjet_axis} one sees that the QCD corrections to the two-jet asymmetry are significantly
    smaller than in the case of the inclusive $b$-jet $\afb$. This makes the two-jet $\afb$ a candidate for a precision observable.
    
    \begin{table}[!h]
\caption{The QCD correction terms  $A_1^{2j}$, $A_2^{2j}$ for the two-jet $A_{\rm FB}$ with respect to the $b$-jet axis  using two jet algorithms with several $y_{cut}$.
 Here unpolarized $e^+e^-$ beams are considered. The numbers for $A_1^{2j}$, $A_2^{2j}$ and their scale variations are given in the unit of $10^{-2}$.}
\label{table:bjet_AswithQQQQ}
\begin{center}
\resizebox{1.0\columnwidth}{!}{
\begin{tabular}{ |c|  c  c  c  c c c| }
\hline 
\multicolumn{1}{ |c }{\multirow{2}{*}{~~$y_{cut}$~~}}                   &   \multicolumn{3}{ |c }{ Flavor $k_T$ }              & \multicolumn{3}{ c| }{ Durham }          \\ \cline{2-7}
	 									&$\sigma_S$ $[\rm pb]$		& $A_1^{2j}$ [$10^{-2}$]		  & $A_2^{2j}$ [$10^{-2}$]		&$\sigma_S$ $[\rm pb]$	& $A_1^{2j}$ [$10^{-2}$] 		& $A_2^{2j}$ [$10^{-2}$]\\
\hline
 	0.01 		&5781.3		& $-0.151^{+0.002}_{+0.001}$ &  $~~0.029^{+0.017}_{-0.011}$		&5984.9		 &   $-0.168^{+0.002}_{+0.001}$    & $~~0.032^{+0.018}_{ -0.012}$    \\
	 0.05		&7666.2		 &$-0.390^{+0.005}_{+0.003}$  & $-0.015^{+0.005}_{-0.002}$ 		&8011.6		 &   $-0.467^{+0.006}_{+0.003}$    &$-0.090^{+0.012}_{-0.016}$  \\
	0.10		&8231.5		 & $-0.571^{+0.007}_{+0.004}$ & $-0.061^{+0.008}_{-0.010}$ 		&8548.1		 &  $~~0.696^{+0.009}_{+0.005}$  &$-0.160^{+0.028}_{-0.038}$ \\
	0.15		&8491.6		 & $-0.705^{+0.009}_{+0.005}$ & $-0.109^{+0.018}_{-0.025}$ 		&8754.7		 & $-0.851^{+0.011}_{+0.006}$ 	 &$-0.222^{+0.041}_{-0.057}$\\
\hline  
\end{tabular}}
\end{center}
\end{table}

The QCD correction to the two-jet $\afb$ becomes smaller in magnitude as $y_{cut}$ becomes smaller,\footnote{This was observed before in \cite{Weinzierl:2006yt}
where the two-jet $\afb$ was computed to NNLO QCD for massless $b$ quarks.} which can be understood from the soft behavior of the amplitude.
 It follows from the fact that by enforcing more stringent two-jet cuts this asymmetry probes the soft region, and the leading QCD soft contributions factorize and largely cancel in the two-jet $\afb$.
 Actually, as the numbers in table~\ref{table:bjet_AswithQQQQ} show, in the case of the flavor-$k_T$ algorithm the magnitude of $A_2^{2j}$ does not fall monotonously with decreasing $y_{cut}$. 
 Yet, we checked that if only the flavor non-singlet contributions are taken into account $|A_2^{2j}|$ decreases with  smaller $y_{cut}$, just as $|A_1^{2j}|$  does.
  
On the other hand, as already mentioned above, the QCD correction terms $A_1$, $A_2$ for the inclusive $b$-jet FB asymmetry increase in magnitude, for both jet algorithms, as $y_{cut}$ decreases, cf. the numbers in table~\ref{table:bjet_axis}.
The NLO correction terms $A_1$ and $A_2$ approach, at very small $y_{cut}$,  the corresponding values determined for the $b$-quark axis (cf. table~\ref{table:quark_axis}). 
In particular, the sequence of numbers for the flavor-$k_T$ algorithm shows this clearly, whereas in the case of the Durham jet algorithm $A_1$ changes significantly for $y_{cut} < 0.05$. 
This is understandable especially with the flavor-$k_T$ algorithm, because for a tiny $y_{cut}$, the difference between the direction of the massive $b$-quark and that of the $b$-jet 
becomes very small for each event as the gluons within the jet are either soft or radiated almost collinear to the massive $b$-quark.

Moreover, by comparing tables~\ref{table:bjet_axis} and~\ref{table:thrust_axis}, it is amusing to notice that the values of $A_1,\, A_2$ of the thrust axis $A_{FB}$, which
was used in the experimental measurements~\cite{ALEPH:2005ab,ALEPH:2010aa}, are very close to those of the inclusive $b$-jet $A_{FB}$ in case of the flavor-$k_T$ algorithm with $y_{cut}=0.01$. 

In principle, one may also determine the direction of the thrust axis using the momenta of the jets  in  eq.~\eqref{eqThrustdef}, rather than those of the partons, for each accepted event with at least one $b$-jet.
The antisymmetric cross section determined in this way remains,  to $\mathcal{O}(\alpha^2_s)$,  the same as the one defined with the thrust axis determined by the parton momenta,
while the symmetric $b$-jet cross section changes. 
This is  because the events without $b$-jets (for instance those where a $b$-quark is combined with an anti-$b$-quark into a jet without $b$-flavor) 
will, by definition,  not be included.
 On the other hand they do not contribute to the antisymmetric cross section. 
For instance, for the  flavor-$k_T$ algorithm and $y_{cut}=0.1$, one obtains $A_1 = -0.0270$ and $A_2 = -0.0066$ when the thrust axis is determined by  the momenta of the jets.
 These correction terms are smaller than the corresponding numbers in table~\ref{table:thrust_axis}.

We conclude the discussion of this section with a short comment on the effect of beam polarization on the $b$-jet forward-backward asymmetries. 
As far as the QCD corrections are concerned, beam polarization affects, as discussed above, only the NNLO terms $A_2^{2j}$ and $A_2$. 
In order to quantify the maximal spread of the NNLO correction terms we use again the ratio defined in eq.~\eqref{eq:rimax}.
In the flavor-$k_T$ algorithm with jet resolution parameter $y_{cut}=0.05$ and $0.1$, we get $|R_2^{\mathrm{max}}|=12\%$ and $10\%$ for $y_{cut}=0.05$ and $0.1$, respectively,
in case of the inclusive $b$-jet asymmetry.  
For the two-jet asymmetry we have $|R_2^{\mathrm{max}}|=54\%$ and $17\%$. These numbers are deceptive, because in absolute terms, the NNLO QCD corrections $A_2^{2j}$ are very small 
and significantly smaller than the QCD uncertainties due to the scale choice.

At this point we add, for completeness, a few remarks concerning  the electroweak corrections to $\bbarb$ production and, in particular, to the $b$-quark asymmetries
 at the $Z$ resonance. An overview of the order $\alpha$ QED corrections (real and virtual initial- and final-state  photonic corrections) and approximations 
  to higher orders is given in \cite{Bohm:1989pb}
  and references contained therein. Their size depends on the experimental set-up, i.e., on the photon cuts applied in an experiment. The most important corrections are the initial-state 
  photonic corrections. A recent higher-order calculation in the logarithmic approximation was made in  \cite{Blumlein:2021jdl}; cf. also the references cited therein.
  
  The purely weak corrections to the $b$-quark asymmetries are virtual corrections; i.e., affect all asymmetries by the same amount. They are model-dependent, that is, their size
   depends on whether their are computed in the Standard Model or some of its extension. The NLO weak SM corrections to the $b$-quark asymmetry 
    have long been known, cf. \cite{Bohm:1989pb} for an overview and 
    references therein. They consist of vertex and box corrections and imaginary parts of propagator corrections. From the tables given in \cite{Bohm:1989pb} one can infer, for the
     top quark mass of eq.~\eqref{eq:parameter}, that the SM weak NLO corrections $\Delta\afb^{\rm weak}$ to the leading order $b$-quark asymmetry 
     $(\afb = \afb^{\rm LO} + \Delta\afb^{\rm weak} + ...)$ is of the order  $\Delta\afb^{\rm weak} \simeq -0.0051$. The full two-loop vertex-type weak SM corrections and the mixed weak-QCD corrections were determined in \cite{Awramik:2004ge,Awramik:2006uz,Awramik:2008gi}.
    The following remark is in order here.  The experimental measurements of the $b$-quark asymmetries (the raw asymmetries) are not used directly in the electroweak fits.
    Instead, a  so-called pseudo-observable is used by subtracting from the respective raw asymmetry some (almost) model independent corrections. These include QED corrections, $\gamma-Z$
     interference, and the QCD corrections \cite{Freitas:2004mn,ALEPH:2005ab,ALEPH:2010aa,Abbaneo:1998xt}. The pseudo-observable obtained in this way (that incorporates the weak corrections)
     is then used in a fit to obtain the effective weak mixing angle. Thus the QCD corrections to the $b$-quark asymmetries determined in this paper are an important ingredient in future analyses of this type.

Finally, we address the  polarized $b$-quark forward-backward asymmetry~\eqref{defalrfb}. As we work to lowest order in the electroweak couplings and, therefore, take only $Z$-boson exchange
 at the $Z$ peak into account, one expects that between $\afblr$ and the corresponding asymmetry for unpolarized beams, $\afb$, the following relation holds \cite{Blondel:1987gp}:
 \begin{equation}
  \label{eq:polb}
   \afb = A_e \afblr = \frac{2 g_{Ve} g_{Ae}}{g_{Ve}^2 + g_{Ae}^2}  \afblr \, ,
 \end{equation}
 where  $g_{Ve}$ and  $g_{Ae}$ are the vector and axial-vector couplings of the electron to the $Z$ boson. Using the value of $\sin^2\theta_W$ listed in eq.~\eqref{eq:parameter} we get $\afb/\afblr = 0.2143$.
 We checked that this relation holds for the $b$-quark axis, thrust axis, and the $b$-jet axis definitions of the asymmetry with various $y_{cut}$; i.e., it is not affected by the QCD corrections.
On the other hand, in the case of top-quark pair production in the high energy limit $s \gg m_Z^2$, the leading order result for this ratio reads  
\begin{equation}
  \label{eq:poltHE}
  \frac{\afb}{\afblr} = \frac{2 g_{Ve} g_{Ae} g_{Vt} + g_{Ae} Q_{e} Q_{t}}{(g_{Ve}^2 + g_{Ae}^2) g_{Vt} + g_{Ve} Q_{e} Q_{t} }\, ,
 \end{equation}
 where $g_{Vt}$ is the vector coupling of the top quark to the Z boson and $Q_{e} $ and $Q_{t}$ are the charges of the electron and top quark, respectively.
 Using the value of $\sin^2\theta_W$ listed in eq.~\eqref{eq:parameter}, we get $\afb/\afblr =3.170$ which sets the high-energy limit for the numbers in table~\ref{ratPolUn}.



\section{Conclusions}
\label{sec:concl}
 We have computed the second-order QCD corrections to the top-quark forward-backward 
 asymmetry in $e^+e^- \to \tbart$  collisions for various c.m.~energies above the $\tbart$
  threshold and to several $b$-quark FB asymmetries at the $Z$ resonance.  These asymmetries should
   play an important role in precision studies, especially in the measurement of the electroweak couplings
    of  heavy quarks at future electron-positron colliders.
  As a new feature we have investigated the effect of $e^+$ and $e^-$ beam polarization.
   We have identified the contributions at NLO and NNLO QCD that are affected by polarized beams. 
We considered a set of benchmark polarizations and found that the relative effects of $e^\pm$ polarizations on the QCD correction factors are, for $\tbart$ production, quite sizeable.
However, in absolute terms, they change the top-quark asymmetry only by an amount of the order of a few per mille, which is comparable to the projected uncertainty of this
observable at future high-energy high-luminosity electron-positron colliders. 
    In the case of $\bbarb$ production at the $Z$ peak and to lowest order in the electroweak couplings
    only the NNLO QCD corrections are affected by beam polarization and the resulting overall effect 
     on the $b$ quark asymmetries is at the per mille level.
     
     Our computational set-up allows also for the calculation of differential distributions, and we 
     demonstrated this by determining the polar angle distribution of the top quark at $\sqrt{s}=500$~GeV.
     We analyzed also, both for $t$ and $b$ quarks, the polarized forward-backward asymmetry that combines data taken with
      two opposite beam polarizations.

       As to $\bbarb$ production at the $Z$ peak we computed, apart for the FB asymmetries with respect to $b$-quark direction and the 
        oriented thrust direction, 
        also an inclusive $b$-jet asymmetry and a two-jet asymmetry, both for the flavor-$k_T$ and the Durham jet clustering 
        algorithm.
       The $b$-jet asymmetries were, to our knowledge, not yet investigated to NNLO QCD for massive $b$ quarks.
       The QCD corrections to the two-jet asymmetry are significantly smaller than those of the other $b$-quark asymmetries.
        This qualifies it as a precision observable for the determination of the neutral-current couplings of $b$ quarks.

\section*{Acknowledgments}

This work was supported by the Natural Science Foundation of China under contract No.12205171, No.12235008.
   
\clearpage 
 

\end{document}